\documentclass[twocolumn]{aastex701}

\def\h{\hskip 0.0 mm}

\def\apjl{{ApJL}}
\def\apjs{{ApJS}}
\def\asec{$^{\prime\prime}$}

\def\e#1{$\times$10$^{#1}$}

\def\farcs{\hbox{$.\mkern-4mu^{\prime\prime}$}}

\def\hal{H$\alpha$}
\def\hb{H$\beta$}
\def\pal{Pa$\alpha$}
\def\pb{Pa$\beta$}

\def\kms{km s$^{-1}$}

\def\lax{{$\mathrel{\hbox{\rlap{\hbox{\lower4pt\hbox{$\sim$}}}\hbox{$<$}}}$}}
\def\gax{{$\mathrel{\hbox{\rlap{\hbox{\lower4pt\hbox{$\sim$}}}\hbox{$>$}}}$}}
\def\simlt{\lower.5ex\hbox{$\; \buildrel < \over \sim \;$}}
\def\simgt{\lower.5ex\hbox{$\; \buildrel > \over \sim \;$}}
\def\lum{erg s$^{-1}$}

\def\micron{{$\mu$m}}

\def\cm2{cm$^{-2}$}

\def\heii{\ion{He}{2}}

\def\sii{[\ion{S}{2}]}

\def\lbol{$L_{{\rm bol}}$}

\shorttitle{Changing-look Phenomenon in NGC 3786}
\shortauthors{Son et al.}

\begin{document}

\title{
The Changing-look Phenomenon Accompanied by an Accretion Mode Transition in NGC 3786
}

\author[0000-0002-5346-0567]{Suyeon Son}
\affiliation{Kavli Institute for Astronomy and Astrophysics, 
Peking University, Beijing 100871, People's Republic of China}
\email[]{}

\author[0000-0002-3560-0781]{Minjin Kim}
\affiliation{Department of Astronomy, Yonsei University, 
50 Yonsei-ro, Seodaemun-gu, Seoul 03722, Republic of Korea}
\email[]{mkim.astro@yonsei.ac.kr}

\author[0000-0001-6947-5846]{Luis C. Ho}
\affiliation{Kavli Institute for Astronomy and Astrophysics, 
Peking University, Beijing 100871, People's Republic of China}
\affiliation{Department of Astronomy, School of Physics, Peking University, Beijing 100871, People's Republic of China}
\email[]{}

\author[0000-0002-6925-4821]{Dohyeong Kim}
\affiliation{Department of Earth Sciences, Pusan National University, Busan 46241, Republic of Korea}
\email[]{}

\author[0000-0002-5857-5136]{Taehyun Kim}
\affiliation{Department of Astronomy and Atmospheric Sciences, 
Kyungpook National University, Daegu 41566, Republic of Korea}
\email[]{}

\correspondingauthor{Minjin Kim}
\email{mkim.astro@yonsei.ac.kr}

\begin{abstract}
To reveal the physical origin of the changing-look (CL) phenomenon in NGC 3786, which transitioned from type 1.8/1.9 to type 1, we present an analysis of long-term spectral monitoring in the optical and near-infrared obtained with Gemini/GMOS-N and Gemini/GNIRS, respectively. Since the onset of the CL phenomenon, NGC 3786 has remained $\sim 1-1.5$ mag brighter in the mid-infrared than in the pre-CL stage, whereas the optical continuum has changed only moderately ($\sim 0.2-0.3$ mag). Spectroscopic analysis further reveals that while the fluxes of the broad Pa$\beta$ and Pa$\alpha$ lines were enhanced over a two-year follow-up period, the flux of the broad H$\alpha$ line remained unchanged. We propose that observed temporal variations in the continuum and line flux ratios disfavor a tidal disruption event origin. Instead, the observations can be primarily explained by a gradual change in line-of-sight extinction driven by variations in the torus covering factor, which is determined by the Eddington ratio and the accretion mode. An additional mechanism, arising from the physical conditions within the broad-line region, may partially account for the temporal evolution of the flux ratios. Our study highlights the importance of investigating the CL phenomenon in intermediate-type active galactic nuclei associated with outbursts detected only in the mid-infrared to explore the detailed structural evolution of nuclear activity.
\end{abstract}


\section{Introduction} 
Supermassive black holes (BHs) are ubiquitous in massive galaxies. While most remain dormant, some are powered by gas accretion, a phenomenon known as an active galactic nucleus (AGN). An AGN consists of various structures, including an accretion disk, a broad-line region (BLR), a dusty torus, and a narrow-line region, that emit radiation across all observed wavelengths. Based on the presence or absence of broad emission lines in the UV and optical wavelengths, AGNs are classified as type 1 or 2, respectively. According to the conventional AGN unification model, AGNs share a common structure, and their types are determined solely by orientation to the line-of-sight, which determines whether the BLR and accretion disk are obstructed by the optically thick dusty torus \citep{antonucci_1993,urry_1995}.

This paradigm is challenged by the discovery of a subset of AGNs that undergo a type transition, referred to as changing-look (CL) AGNs \citep{pastoriza_1970,khachikian_1971,ricci_2023b}. While the majority of CL AGN transitions discovered to date are thought to be driven by intrinsic changes in the AGN luminosity, a smaller subset of the population may be attributed to variations in line-of-sight obscuration \cite[e.g.,][]{zeltin_2022}. In addition, previous studies suggest that at least some CLs are associated with tidal disruption events (TDEs; e.g., \citealt{merloni_2015,trakhtenbrot_2019,li_2024}).  
Notably, some objects exhibiting repeating phenomena \citep{marin_2019,wang_2024b,wang_2025,kollatschny_2026arXiv} on short timescales ($\sim$ a few years) highlight the need to refine the unified paradigm beyond a simple orientation-based interpretation. 

CL AGNs are often accompanied by extreme brightness changes in the optical continuum, so they have been identified using optical continuum variability (e.g., \citealt{yang_2018,macleod_2019,wang_2026}). Therefore, the currently available sample may be biased toward the extreme end of the CL phenomenon. To fully understand the CL process, observing moderate CL AGNs or the early stages of CL phenomena can be beneficial. Those missing populations can be detected, possibly in intermediate-type AGNs between type 1 and type 2, where broad \hal\ line is clearly detected while broad \hb\ line is absent or very weak \cite[e.g.,][]{osterbrock_1976, ho_1997b}. Furthermore, those objects can be identified through infrared observations, as they can occur in AGNs with significant obscuration. CL phenomena in the MIR are vital to this end.       

We serendipitously discovered a mid-infrared (MIR)-only outburst in NGC 3786, which may have begun around 2020, by comparing the MIR light curve from the Near-Earth Object Wide-field Infrared Survey Explorer (NEOWISE; \citealt{mainzer_2011}) with the optical light curve from the Zwicky Transient Facility (ZTF; \citealt{bellm_2019}). Before the MIR eruption, NGC 3786 had been classified as a type 1.8 or 1.9 AGN \citep{goodrich_1983}.  Through our first optical and near-IR (NIR) spectroscopic follow-up with Gemini in 2022, we showed that AGN type changed from 1.8/1.9 to 1 by confirming that broad \hb\ and \pb\ lines newly appeared and broad \hal\ line significantly enhanced after the outburst began \citep{son_2022b}. However, it was difficult to distinguish whether the CL phenomenon was driven by a change in line-of-sight obscuration or by an intrinsic increase in the accretion rate, possibly due to a TDE. In this paper, we aim to clarify the origin of the MIR-only eruption and to track BLR evolution after the outburst by analyzing two additional optical and NIR spectroscopic follow-up observations obtained with consistent instrumental settings with those used in our first follow-up observation.

\section{Brightness Change in Broadband Light Curves}

To examine continuum variability since the study from \cite{son_2022b}, we compiled broadband monitoring data in the optical and MIR from ZTF ($g$ and $r$ bands) and NEOWISE (W1 and W2 bands), respectively. We averaged ZTF and NEOWISE light curves in 6-month bins, corresponding to the typical NEOWISE cadence, after applying $3\sigma$ clipping in each bin containing more than 3 epochs (Fig.~1; \citealp{son_2022, son_2023b}). For ZTF light curves, we adopted the standard deviation of photometric measurements in the MJD bin as the magnitude uncertainty. We calculated magnitude uncertainties for binned NEOWISE light curves using Eq.~(2) of \cite{lyu_2019}, which is based on the photometric stability of NEOWISE as quantified with calibration stars. 

As shown in Figure~1, NGC 3786 has brightened by an average of 0.68 and 1.02 mag in the W1 and W2 bands, respectively, compared to the pre-outburst state (MJD $\lesssim 59000$). In contrast, it brightened on average by only 0.11 mag in $g$ and 0.14 mag in $r$. Additionally, the maximum brightness differences before and after the outburst are 0.27, 0.25, 1.07, and 1.53 mag in the $g$, $r$, W1, and W2 bands, respectively. Notably, during this period, the W1$-$W2 color changed significantly (on average $\sim0.21$ and 0.55 mag for before and after the MIR outburst), but no clear evidence of the change in $g-r$ color was detected. Overall, the outburst in NGC 3786 is significantly stronger in the MIR than in the optical bands, suggesting that it may occur in a heavily obscured region. 

Non-detection of significant variations in the optical band may be attributed to a combination of heavy obscuration and contamination from the dominant host galaxy. The mean $g-r$ color of $\sim 0.8$ mag derived from ZTF measurements is notably redder than typical low-redshift QSOs \citep[e.g.,][]{richards_2002}, suggesting that the optical flux is indeed dominated by stellar light from the host. To provide a quantitative estimate, we consider a scenario where an outburst increases the optical flux by 1.0 mag, a value inferred from the observed magnitude change in the W2 band. This remains a conservative assumption, as variability amplitudes in the optical regime are typically much larger than those observed in the infrared \citep[e.g.,][]{kim_2024}. Given the observed $g$-band mean magnitude of $\sim 14.9$ mag and an RMS variability of $\sim 0.1$ mag, we calculate the expected magnitude of the underlying nucleus. Our results indicate that the nucleus must be fainter than 18.4 mag in the $g$-band to remain undetected within the observed noise levels. This limit is fainter than the expected optical luminosity, inferred from the bolometric luminosity ($L_{\text{bol}} \approx 10^{43.2} \text{--} 10^{44.6}$ erg s$^{-1}$), derived from the \hal\ flux by approximately a factor of $10-100$ (see \S{4}). This discrepancy strongly suggests that the lack of optical continuum variability is primarily due to significant obscuration rather than host galaxy contamination alone.

More importantly, we fail to detect a decline in brightness after the outburst, indicating it is unlikely to originate from a TDE or a flare-like eruption. However, the decline time scale can vary significantly due to the complex dust structure surrounding the nucleus. Therefore, we cannot completely rule out the possibility of the TDE solely based on the continuum variability.

\begin{figure}[t!]
\centering
\includegraphics[width=0.5\textwidth]{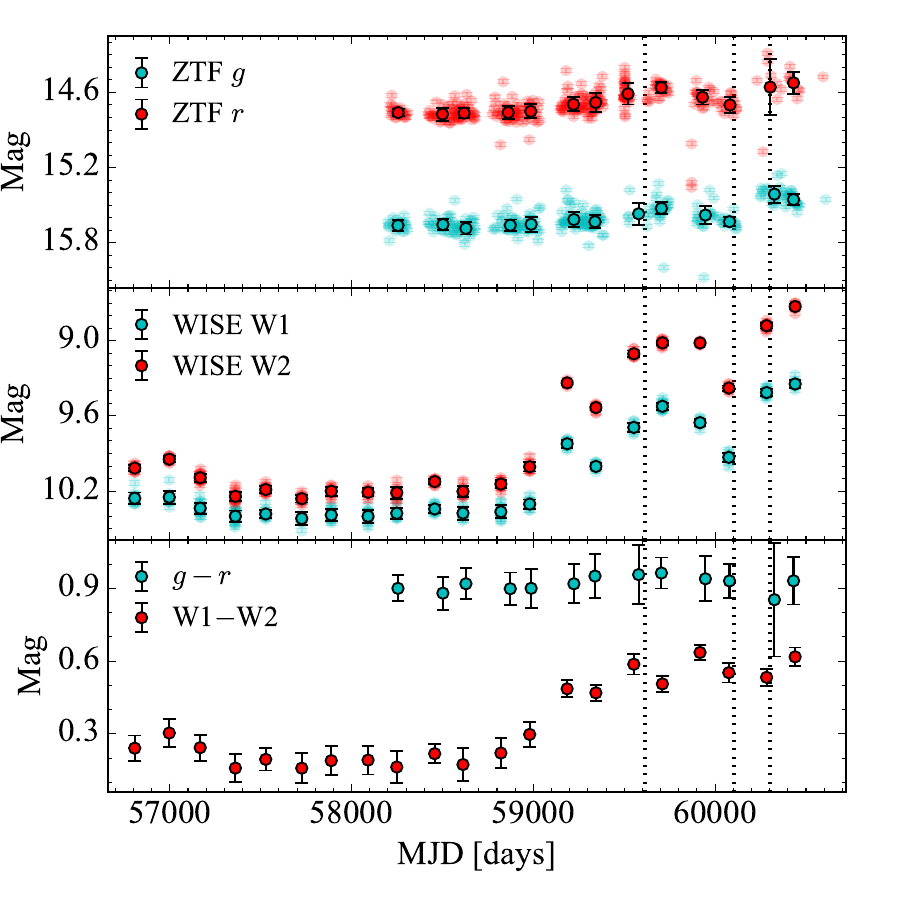}
\caption{
Light curves of NGC 3786 in the ZTF $g$ and $r$ band (top) and the WISE W1 and W2 band (middle). 
The error bars represent the $1\sigma$ uncertainty. In the bottom panel, the $g-r$ and W1$-$W2 colors are represented. Black vertical dotted lines indicate the epochs when the follow-up spectra were obtained with the Gemini telescope.
\label{fig:fig1}}
\end{figure}

\section{Spectroscopic Follow-up Observations}

\subsection{Observation and Data}

To reveal the nature of the MIR-only outburst, we carried out spectral monitoring of NGC 3786 with constant instrumental settings as in our previous spectroscopic observation \citep{son_2022b}. To monitor broad emission lines in the optical (\heii, \hb, and \hal) and the NIR (\pb\ and \pal), we used GMOS-N and GNIRS at the Gemini telescope. We obtained GMOS-N and GNIRS spectra on 2022-02-02, 2023-06-10, 2023-11-06/04 (GMOS-N/GNIRS), and 2023-12-23. We excluded the spectral data from the third observation (2023-11-06/04) because the analyzed spectra had severe artifacts. The airmass ranged from $ 1.1$ to $1.2$, and the seeing was between $ 0\farcs75$ and $1\farcs00$ throughout the observing dates.

For the GMOS-N observation, we obtained two spectra with exposure times of 280 seconds per epoch, using the B600 and B480 gratings with a $0\farcs75$ slit. We attempted to maintain the same settings (e.g., position angle and slit width) as in \cite{son_2022b} to facilitate a direct comparison of observations across different epochs. However, we replaced the GMOS-N B600 grating with the B480 grating due to significant degradation of the former. The resulting change in spectral resolution (from $R \approx 1100$ to $1000$) was accounted for during spectral analysis and had a minimal impact on our results. To mitigate chip gaps via two-point spectral dithering, the central wavelengths were set to 0.520 and 0.525--0.527 $\mu$m for the two exposures.

For GNIRS spectra, the cross-dispersed mode was adopted to provide a wide wavelength coverage ($0.8-2.5$ \micron) to encompass broad Paschen lines and the continuum. Using a 32~lines mm$^{-1}$ grating and a 0\farcs3 slit to achieve a spectral resolution of 1800, we took 12 spectra per epoch, including 4 for blank sky, with exposure times of 60 seconds because the $7\arcsec$ slit length of the short camera is insufficient to cover the sky simultaneously. More detailed descriptions of observational settings for GNIRS are available in \cite{son_2022b}.


We performed data reduction using the \texttt{Gemini/GMOS} and \texttt{Gemini/GNIRS} IRAF packages for optical and NIR spectra, respectively, following the same procedures as described in \cite{son_2022b}. For GMOS-N spectra, we first processed them following standard procedures: bias subtraction, flat-field correction, and cosmic-ray removal. We performed wavelength calibration using arc images and sky subtraction using blank-sky spectra obtained from the same slit of the science target. We extracted 1D spectra from a 0\farcs75$\times$5\asec\ aperture. Flux calibration was applied using spectra of standard stars Feige 66, EG 131, and Wolf 1346 for the first, second, and last follow-up observations, respectively. For GNIRS spectra, we performed flat-fielding using two flat images obtained with different ramps. This ensured a high signal-to-noise ratio in both low- and high-order spectra, enabling reliable recovery of response functions across the entire wavelength range. Additionally, we applied distortion correction with pinhole spectra, wavelength calibration using arc images obtained with an argon lamp, sky subtraction, and extraction of 1D spectra with a 3 \asec\ diameter. Flux calibration and telluric correction were carried out using the standard star HIP 57239 for all epochs following \citet{vacca_2003}.

\begin{figure*}[htp]
\centering
\includegraphics[width=0.9\textwidth]{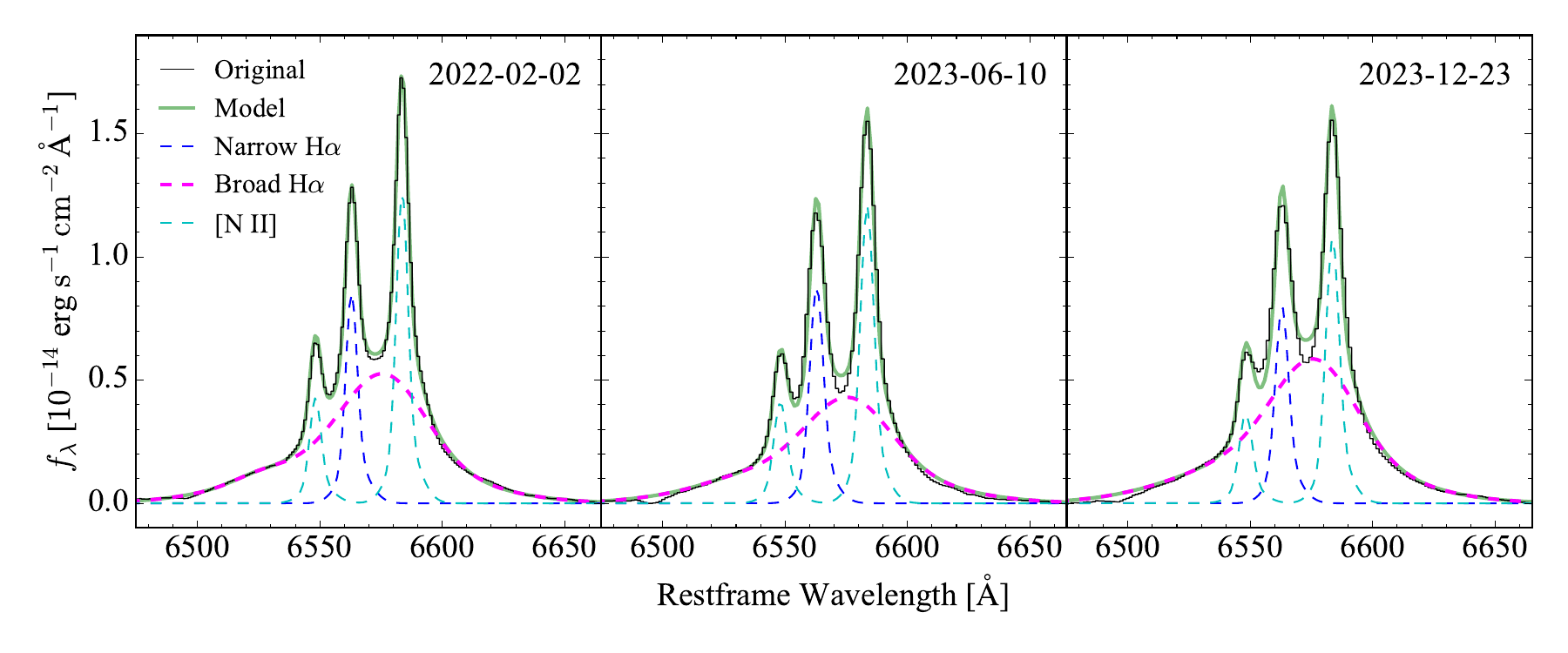}
\includegraphics[width=0.9\textwidth]{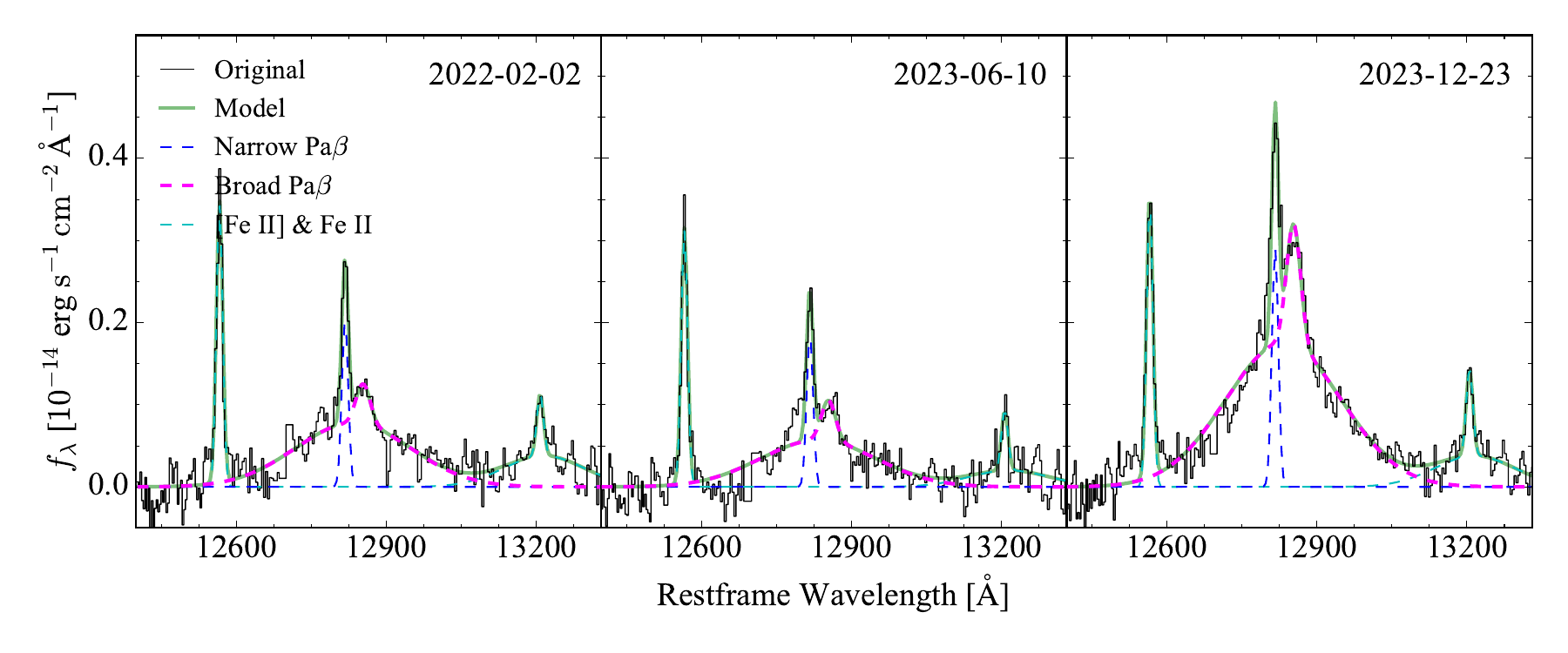}
\includegraphics[width=0.9\textwidth]{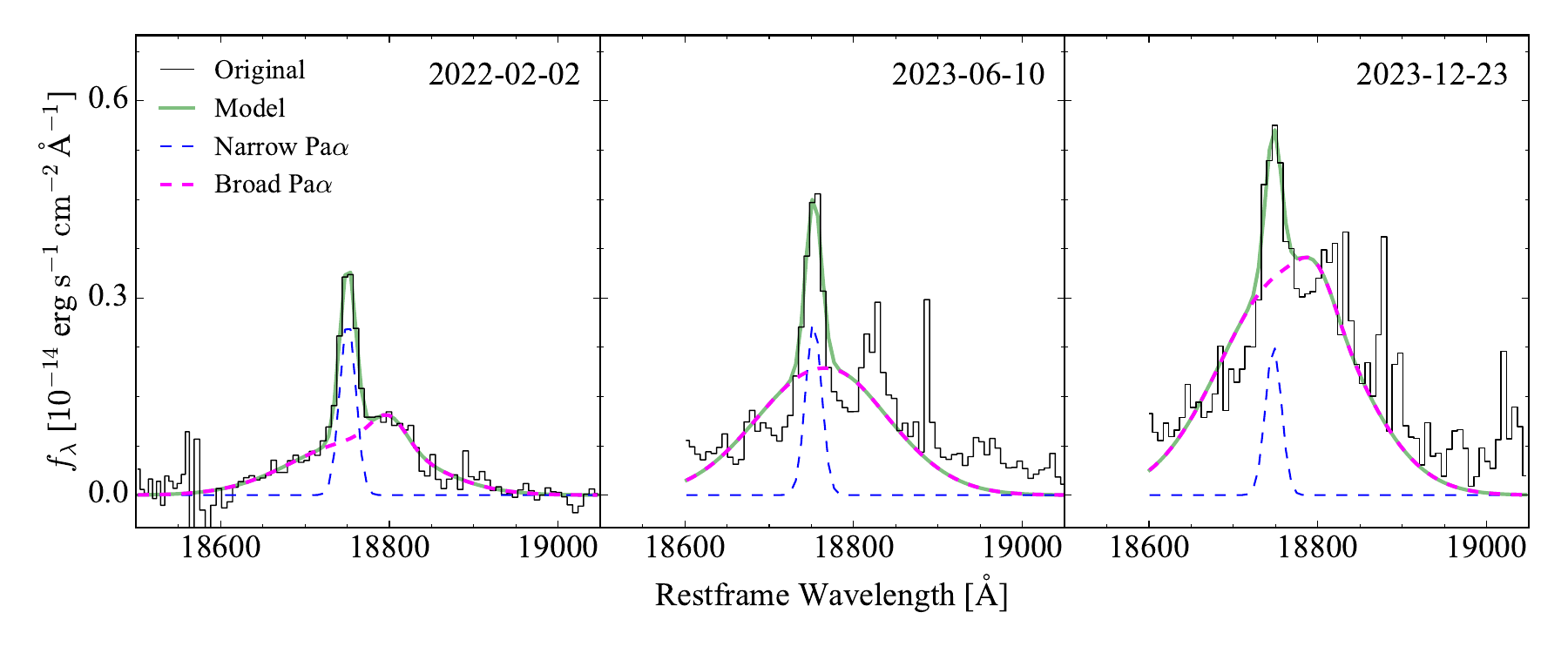}
\caption{
Fitting results for \hal\ (top), \pb\ (middle), and \pal\ (bottom) regions. During the fitting procedure, the profiles of broad emission lines are fixed to the first observation. In all panels, the black and green lines represent the original spectra and best-fit model, respectively, while the blue and magenta dashed lines denote the best-fit models for the narrow and broad components of \hal\ (top), \pb\ (middle), and \pal\ (bottom) emission lines, respectively. The cyan line represents the other emission lines in the fitting region. The spectrum at shorter wavelengths ($\lambda < 18600$ \AA) in the Pa$\alpha$ region was excluded from the fit, as it is heavily affected by telluric absorption.
\label{fig:fig2}}
\end{figure*}

\begin{deluxetable*}{lcrrrr}
\tablecolumns{13}
\tablenum{1}
\tablewidth{0pc}
\tablecaption{Spectral Properties of NGC 3786 \label{tab:table1}}
\tablehead{
\colhead{\h Line} &
\colhead{\h Fit type} &
\colhead{\h Flux} &
\colhead{\h FWHM} &
\colhead{\h Center} &
\colhead{\h Velocity shift} \\
\colhead{\h (1)} &
\colhead{\h (2)} &
\colhead{\h (3)} &
\colhead{\h (4)} &
\colhead{\h (5)} &
\colhead{\h (6)}
}
\startdata
Narrow \hal &\h fix    &\h 6.44/7.20/6.56   &\h 219/186/191 &\h 6563.1/6562.5/6563.3\\ 
         \h &\h free &\h 6.44/6.51/6.07   &\h 219/186/191 &\h 6563.1/6562.5/6563.3\\ 
\hline
Broad \hal  &\h fix           &\h 32.7/27.0/35.7  &\h 2240/2260/2205 &\h 6576.2/6576.2/6575.8 &\h +598/+625/+570\\ 
         \h &\h fix$^{\rm s}$ &\h 13.5/11.4/16.0  &\h 1661/1655/1661 &\h 6578.0/6578.0/6578.0 &\h +680/+707/+670\\
         \h &\h fix$^{\rm s}$ &\h 3.7/4.9/5.5  &\h 2593/2589/2593 &\h 6548.7/6548.7/6548.7 &\h $-660$/$-632$/$-669$\\
         \h &\h fix$^{\rm s}$ &\h  15.5/10.7/14.2  &\h 4224/4222/4224 &\h 6567.5/6567.5/6567.5 &\h +201/+228/+192\\
         \h &\h free        &\h 32.7/26.3/35.3  &\h 2240/2378/2377 &\h 6576.2/6565.0/6573.3 &\h +598/+114/+456\\ 
         \h &\h free$^{\rm s}$ &\h 13.5/3.1/2.0  &\h 1661/834/1252 &\h 6578.0/6559.7/6522.8 &\h +680/$-128$/$-1863$\\ 
         \h &\h free$^{\rm s}$ &\h 3.7/5.1/27.3  &\h 2593/1791/2206 &\h 6548.7/6533.3/6573.1 &\h $-660$/$-1341$/+447\\ 
         \h &\h free$^{\rm s}$ &\h  15.5/18.0/5.9  &\h 4224/1941/4498 &\h 6567.5/6580.0/6580.0 &\h +201/+798/+761\\
\hline
Narrow \pb  &\h fix &\h 3.3/3.0/4.7  &\h 284/284/284 &\h 12817.5/12816.7/12816.8\\ 
        \h  &\h free &\h 3.3/2.0/2.7  &\h 284/284/284 &\h 12817.5/12817.0/12816.2\\ 
\hline
Broad \pb   &\h fix &\h 24.2/18.2/56.0  &\h 3540/2352/2445 &\h 12851.4/12853.9/12852.4 &\h +791/+868/+831\\ 
        \h  &\h fix$^{\rm s}$ &\h 1.7/1.8/5.2  &\h 747/747/747 &\h 12853.7/12853.9/12853.6 &\h +845/+868/+859\\ 
        \h  &\h fix$^{\rm s}$ &\h 22.4/16.5/50.7  &\h 6356/6355/6355 &\h 12833.2/12833.4/12833.4 &\h +367/+390/+388\\ 
        \h  &\h free &\h 24.2/17.2/60.0  &\h 3540/3181/2996 &\h 12851.4/12822.7/12836.9 &\h +791/+133/+484\\ 
        \h  &\h free$^{\rm s}$ &\h 1.7/10.1/17.8  &\h 747/2647/2139 &\h 12853.7/12815.4/12837.8 &\h +845/$-37$/+505\\ 
        \h  &\h free$^{\rm s}$ &\h 22.4/7.1/42.0  &\h 6355/5102/8027 &\h 12833.2/12900.0/12840.7 &\h +367/+1930/+572\\
\hline
Narrow \pal &\h fix &\h 6.4/6.3/5.4  &\h 284/284/284 &\h 18750.7/18752.7/18747.5\\ 
         \h &\h free &\h 6.4/7.8/7.3  &\h 284/284/284 &\h 18750.7/18752.5/18747.6\\ 
\hline
Broad \pal  &\h fix &\h 18.6/37.8/68.1  &\h 2168/2887/2775 &\h 18792.9/18764.0/18787.9 &\h +674/+181/+645 \\
         \h &\h fix$^{\rm s}$ &\h 2.5/0.2/2.7  &\h 766/766/766 &\h 18799.5/18799.7/18799.7 &\h +779/+750/+833 \\
         \h &\h fix$^{\rm s}$ &\h 16.0/37.6/65.4  &\h 2924/2924/2924 &\h 18762.9/18763.1/18763.1 &\h +195/+166/+249 \\
         \h &\h free &\h 18.6/42.5/76.7  &\h 2168/2052/3783 &\h 18792.9/18822.1/18813.8 &\h +674/+1109/+1056 \\
         \h &\h free$^{\rm s}$&\h 2.5/2.4/2.8  &\h 766/241/451 &\h 18799.5/18824.9/18815.5 &\h +779/+1154/+1083 \\
         \h &\h free$^{\rm s}$&\h 16.0/37.6/65.4  &\h 2924/3753/3755&\h 18762.9/18795.0/18785.3 &\h +195/+678/+602 \\
\enddata
\tablecomments{
Col. (1): Emission line. 
Col. (2): Fit type: ``fix'' denotes fitting with fixed velocity shifts; ``free'' denotes unconstrained fitting.
Col. (3): Line flux in units of $10^{-14}$ erg s$^{-1}$ cm$^-2$. In columns (3) through (6), measurements for the various epochs are delimited by `/', listed in chronological order.
Col. (4): FWHM of the emission line in units of \kms.
Col. (5): Rest-frame central wavelength of the emission line in units of \AA.
Col. (6): Velocity shift of the broad emission line relative to the narrow emission line in units of \kms. \\
$^{\rm s}$ Measurements of each Gaussian component.
}
\end{deluxetable*}

\subsection{Spectral Fitting}
To examine variations in the flux and kinematics of the broad emission lines, we performed spectral fits to three optical and three NIR spectra obtained with consistent instrumental settings over a two-year baseline. The fitting was restricted to the \hal, \pb, and \pal\ regions, as the broad \heii\ line was not detected, and the presence of the broad \hb\ line could not be confirmed in the second and third epochs due to low S/N in the \hb\ spectral region.

Before line fitting, we first subtracted the local continuum, which was modeled with a first-order polynomial (e.g., \citealt{denney_2009}). For the \hal\ line from the GMOS-N spectra, we modeled the narrow-line profiles using the \sii\ $\lambda\lambda$6716, 6731 doublet as a template, with each \sii\ line represented by two Gaussian components. For the GNIRS \pal\ and \pb\ lines, where narrow and broad components were clearly distinguishable, we fitted the narrow and broad emission lines simultaneously, using a single Gaussian for each narrow component. For the broad components of \hal, \pb, and \pal, we iteratively added Gaussian components as needed. Overall, the profiles were typically well-represented by a double Gaussian, while broad \hal\ required three Gaussians. Finally, the flux was recalibrated by normalizing the spectra, assuming that the narrow emission-line flux remained constant throughout our monitoring period.

To investigate whether the central velocity of the broad emission lines shifted over time, we performed spectral fitting of the data from the final two epochs using two approaches: (1) fixing the velocity shifts of the broad Gaussian components to the values from the first epoch, and (2) allowing them to vary freely. Based on this experiment, we found that the fitting results were consistent regardless of the method adopted, suggesting that the velocity shift between epochs is negligible. Therefore, throughout this paper, we adopt the fitting results obtained with fixed velocity shifts. Broad \hal, \pb, and \pal\ lines were clearly detected in all epochs. Notably, the broad Paschen lines brightened gradually and significantly by a factor of $\sim2.3-3.7$, while the flux of the broad \hal\ line remained unchanged within a $\sim20\%$ (Fig.~2 and Tab.~1). Table 1 summarizes the spectral properties of different epochs. For the broad \hal\ emissions, the typical uncertainties are $\sim0.03$ dex and $\sim 30$ \kms, for the flux and central wavelength, respectively. For the \pal\ lines, the typical uncertainties in the flux and central wavelength are $\sim 0.1$ dex and $\sim 200$ \kms, respectively. The possible implications of these findings are presented in \S{5}.

\section{AGN properties}

To examine the physical origin of the CL phenomenon associated with the MIR outburst, it is vital to reliably measure the evolution of AGN properties during this phase. The BH mass can be estimated using various methods depending on the AGN type. One approach is the virial estimator, expressed as $M_{\rm BH} \approx f \frac{r v^2}{G}$, where $v$ and $r$ represent the velocity dispersion and size of the BLR, respectively, and $f$ is the virial factor determined by the BLR kinematics and structure \cite[e.g.,][]{peterson_1999}. Utilizing the BLR size-luminosity relation, the BH mass can be readily estimated from a single-epoch spectrum for typical type 1 AGNs \cite[e.g.,][]{kaspi_2000,bentz_2013}. However, this method cannot be applied to our target, as it remains uncertain whether the BLR is fully virialized in this outburst stage. For example, 1ES 1927+654, associated with a CL phenomenon potentially triggered by a TDE, exhibits dramatic evolution in BLR kinematics, suggesting that the BLR is not virialized \citep{li_2022}. 

Instead, \cite{son_2022b} estimated the BH mass based on the relation between the BH mass and the bulge mass ($M_{\rm bulge}$). To achieve this, they performed a 2D image decomposition of the 3.6 $\mu$m data from the Infrared Array Camera (IRAC) on the {\it Spitzer} Space Telescope, yielding a bulge magnitude of $-18.72$ mag. By adopting the mass-to-light ratio from \cite{munoz_2013} and the $M_{\rm BH}$–$M_{\rm bulge}$ relation from \cite{kormendy_2013}, \cite{son_2022b} derived a BH mass of $\log\ (M_{\rm BH}/M_{\odot})=6.70$, which we adopt for the subsequent analysis in this study. The uncertainty in $M_{\rm BH}$ is dominated by the intrinsic scatter ($0.3$–$0.5$ dex) in the $M_{\rm BH}$–$M_{\rm bulge}$ relation of inactive galaxies, depending on the bulge classification, rather than the measurement error of the bulge magnitude ($\sim0.1$ mag).

The AGN bolometric luminosity (\lbol) can be estimated using various proxies, including the continuum luminosity in the optical or MIR bands and the line luminosities of narrow or broad emissions. For our target, the optical continuum cannot be reliably measured due to significant obscuration. Instead, we computed the bolometric luminosity from the broad \hal\ luminosity \citep{greene_2005,richards_2006}, which exhibits a tight, nearly linear correlation with the optical continuum. We utilized the broad \hal\ line because it was clearly detected in all epochs. Furthermore, emission lines from the BLR respond to ionizing continuum variations more rapidly than the MIR continuum or narrow lines. We converted the broad \hal\ luminosity to the continuum luminosity at 5100 \AA\ ($L_{5100}$) using the relation from \cite{greene_2005}, and then converted $L_{5100}$ to \lbol\ using the bolometric correction from \cite{richards_2006}. We derived $\log\ (L_{\rm bol}/{\rm erg\ s^{-1}})=  
43.31, 43.24$, and $43.34$ for the first, second, and third observations, respectively, without applying extinction corrections.

To account for dust extinction in the BLR, we estimated $E(B-V)$ by comparing the observed broad \hal/\pal\ ratio to the intrinsic value of $\sim$9.00 \citep{kim_2010}, assuming the observed ratio is entirely determined by dust extinction. We adopted the extinction law from \cite{fitzpatrick_1999}. Notably, we found that $E(B-V)$ gradually increased throughout our spectroscopic monitoring, yielding values of 0.92, 1.42, and 1.60 mag. The extinction-corrected bolometric luminosities are 44.05, 44.39, and 44.64 \lum\ for the first, second, and third observations, respectively. The uncertainty in the extinction-corrected bolometric luminosities is approximately 0.1 dex, which is mainly attributed to the low S/N of the broad \pal\ emissions.  

The Eddington ratio was computed as $\lambda_{\rm Edd} \equiv L_{\rm bol}/L_{\rm Edd}$, where $L_{\rm Edd}$ is the Eddington luminosity, using the derived BH mass and AGN bolometric luminosities. Figure 3 shows the Eddington ratio before and after correcting for dust extinction. We adopted the pre-outburst Eddington ratio derived by \cite{son_2022b} from optical spectra obtained with the Perkins 1.8 m telescope \citep{koss_2017}. Note that, for the pre-outburst epoch, extinction correction could not be taken into account reliably, as other broad lines were barely detected or their fluxes could not be estimated robustly from the pre-outburst spectra. Nevertheless, a dramatic rise in the Eddington ratio is clearly evident, temporally linked to the MIR outburst.

\begin{figure}[ht!]
\centering
\includegraphics[width=0.45\textwidth]{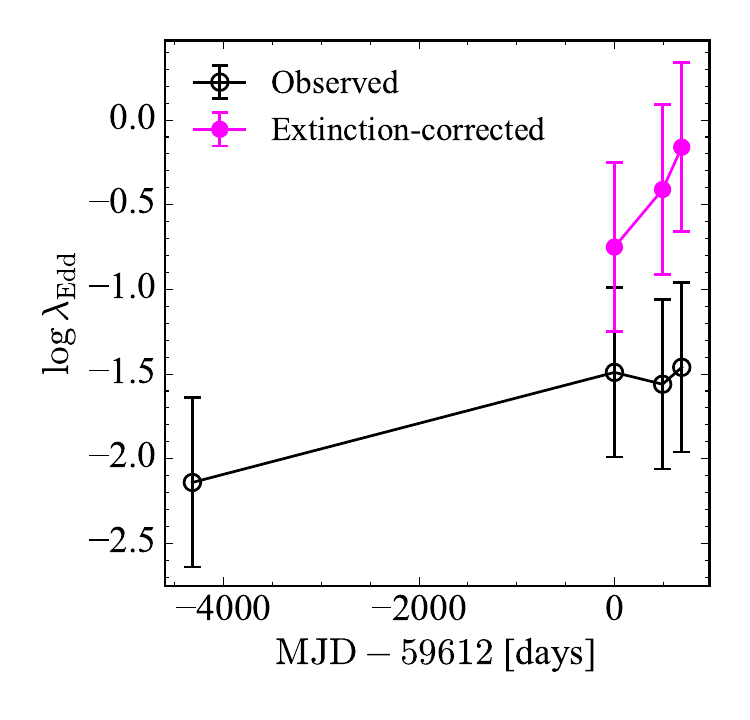}
\caption{
Eddington ratios estimated without (open black circle) and with (filled magenta circle) dust extinction correction. $E(B-V)$ was derived from the observed flux ratio of \hal/\pal. However, this correction could not be applied to the pre-outburst epoch because the broad \pal\ was not detected. The uncertainty in the Eddington ratio is dominated by the systematic scatter in the BH mass estimation.}
\label{fig:fig3}
\end{figure}

\section{Physical Origins of the Outburst}

NGC 3786 exhibited CL behavior, transitioning from type 1.8/1.9 to type 1, likely due to an extreme change in accretion rate, obscured by the abundant dust. Such an extreme accretion change can be caused by a TDE, in which temporal variations of BLR kinematics are expected (e.g., \citealt{merloni_2015,holoien_2016,li_2022}). However, no clear change in the velocity shift of broad emission lines during our spectroscopic follow-up was detected, which reduces the possibility that the CL phenomenon was caused by a TDE. Furthermore, the broad \heii\ line, one of the characteristics of TDE, is not visible in our spectra obtained roughly 400–1300 days after the onset of the MIR outburst. Moreover, the temporal evolution of the MIR light curve differs from that expected for a TDE. If the MIR outburst was caused by an obscured TDE, thermal dust reprocessing of TDE emission is expected to produce a transient-like MIR emission (see the review by \citealt{vanvelzen_2021}). In contrast, the MIR continuum did not decline after the outburst, showing a sustained transition to a higher state instead. Therefore, our results disfavor a TDE origin for the extreme accretion change, suggesting it is likely driven by an intrinsic fluctuation in the accretion rate. However, because the nuclear region can be heavily obscured by significant dust and the target galaxy is undergoing an active merger \citep{derobertis_1998}, the predictions derived from typical optically selected TDEs may not be directly applicable to this phenomenon. Therefore, a TDE origin cannot be ruled out entirely.

Interestingly, after the CL behavior, NGC 3786 exhibited extreme flux variations in the broad Paschen lines and the MIR continuum. However, such variations were not detected in the broad \hal\ line or the optical continuum. In the following sections, we discuss the implications for the observed evolution of the continuum and the BLR in NGC 3786 by examining two plausible scenarios: variations in dust extinction and intrinsic changes in BLR conditions.

\subsection{Variations in Dust Extinction}

In general, the reddening-consistency test for the pure obscuration scenario assumes that the accretion disk and the BLR share a common extinction \citep[e.g.,][]{lamassa_2015}. However, this approach is not feasible for the current study as the optical continuum is not securely detected due to its faintness and significant host galaxy contamination. Furthermore, the broad \hb\ line is undetected in the later epochs, preventing an extinction estimate via the Balmer decrement. Finally, since an $E(B-V)$ derived solely from \hal/\pal\ does not provide a fully independent assessment, we conclude that the pure obscuration scenario cannot be robustly tested with the current dataset. 

The observed CL behavior can be caused by a geometrical change in the dusty torus: a decrease in line-of-sight extinction toward the BLR, driven by a declining torus covering factor. The torus covering factor has been reported to depend on AGN luminosity (e.g., \citealt{maiolino_2007,mor_2011,trefoloni_2025}, but see \citealt{netzer_2016,stalevski_2016}) and/or Eddington ratio (e.g., \citealt{ricci_2017c,zhuang_2018,ricci_2023}). \cite{zhuang_2018} showed that the torus covering factor decreases with increasing Eddington ratio over $\log \lambda_{\rm Edd}\approx-2.0$ to $-0.25$, but reverses (i.e., increases) with increasing Eddington ratio over $\log \lambda_{\rm Edd}\approx-0.25$ to 0.5. These two opposite trends can be explained by two mechanisms. For AGNs with a standard disk \citep{shakura_1973,frank_2002}, which exists at $-2 \lesssim \log \lambda_{\rm Edd} \lesssim -0.5$, an increase in dust sublimation radius reduces the torus covering factor as the AGN brightens (e.g., \citealt{lawrence_1991,simpson_2005,honig_2007}). For AGNs with a high Eddington ratio ($\log \lambda_{\rm Edd} \gtrsim -0.5$), where the accretion flow turns into a slim disk \citep{begelman_1978,abramowicz_1988,wang_2014}, the torus covering factor increases due to radiation pressure driven by the collimated accretion disk radiation perpendicular to the plane of the torus. We note, however, that this interpretation should be treated with caution, as the Eddington ratio estimate is subject to considerable uncertainty ($\sim 0.3-0.5$ dex; \citealp{kormendy_2013}), primarily due to intrinsic scatter in the $M_{\rm BH}$–$M_{\rm bulge}$ relation of inactive galaxies. We therefore conservatively assume a 0.5 dex uncertainty in the BH mass estimate and Eddington ratio. We stress that this uncertainty is systematic rather than random. Consequently, the overall trend in the Eddington ratio remains unchanged.

The accretion disk of NGC 3786 remained in the standard disk regime with $\log \lambda_{\rm Edd}=-0.75$ for the first spectral observation following the MIR outburst, and subsequently transitioned to the slim disk regime with $\log \lambda_{\rm Edd}=-0.41$ and $-0.16$ for the second and third observations (Fig.~3). We cannot compute the extinction-corrected Eddington ratio before the outburst, but the MIR light curve, which is relatively free from dust extinction, shows that NGC 3786 was in a lower state before MJD $\lesssim 59000$, making it likely that it was in the standard disk regime. The MIR outburst observed after MJD $\sim59000$ may have decreased the torus covering factor, leading to reduced dust extinction and a subsequent AGN type change from 1.8/1.9 to 1. Following the first follow-up observation, the collimated radiation of a slim disk likely led to an increase in the torus covering factor and, consequently, greater dust extinction. A simultaneous increase in dust extinction alongside increasing AGN brightness can result in only a slight increase in the optical continuum and broad \hal\ line brightness. In contrast, the broad Paschen lines and the MIR continuum show significant variability because the optical depth effect is substantially lower than for the broad \hal\ line and optical continuum. 

In summary, we have witnessed a structural change in the dusty torus of NGC 3786 associated with the CL phenomenon. This demonstrates that CL AGNs occurring in intermediate-type AGNs, or those associated with MIR-only outbursts, are essential for studying the detailed structural evolution of AGNs.

\subsection{Intrinsic Changes in BLR Conditions}

A dramatic change in the flux ratio between broad emission lines can also be caused by physical conditions in the BLR, which can affect the radiation mechanism, rather than the dust obscuration \cite[e.g.,][]{netzer_1975,davidson_1979,wu_2023,son_2025b}. Therefore, this possibility needs to be further explored to explain the observed trend in NGC 3786. An increased accretion rate can drive stronger disk winds (i.e., larger mass outflows), which can lead to the formation and development of dense gas in the BLR \cite[e.g.,][]{li_2022}. This can be an important mechanism for the appearance of broad emission lines in CL AGNs. An outburst in the MIR light curve indicates that NGC 3786 experienced a significant increase in accretion rate and has remained in a high state, which may play a role in the formation of the BLR. 

For AGNs with high Eddington ratios, broad emission lines are enhanced by collisional excitation and consequently become less sensitive to continuum variations. Since the collisional contribution to excitation is more significant for transitions involving lower energy levels \citep{parker_1964}, it affects \hal\ more profoundly than \pb\ and \pal. From this perspective, the broad \hal\ line likely became less sensitive to continuum variations after the first follow-up observation due to this substantial collisional contribution. In contrast, broad Paschen lines are primarily determined by recombination and thus respond more sensitively to continuum fluctuations. While this suggests that changes in the physical conditions of the BLR could drive the observed flux ratio trends, this model alone struggles to explain why the optical continuum remains relatively constant while the MIR continuum varies significantly. Therefore, internal BLR changes cannot entirely account for the observations; instead, it is likely that both mechanisms, structural changes in the dusty torus and variations in line responsivity, contribute to the observed trends.

\section{Conclusion}
To explore the physical origin of the CL phenomenon occurring in NGC 3786 associated with the MIR-only outburst, we carried out the long-term optical and NIR spectroscopic monitoring using GMOS/GNIRS over $\sim 2$ years. In combination with the optical and MIR continuum variability obtained from ZTF and NEOWISE, we reached the following conclusions.

\begin{itemize}
    \item While the nucleus of NGC 3786 remains bright in the MIR, without any signature of decline since the onset of the outburst, the optical continuum exhibits significantly less variability than the MIR continuum. The persistence of the MIR outburst throughout our follow-up observations suggests that this phenomenon is not driven by a transient event, such as a TDE, which would exhibit a decaying light curve.
    
    \item During the spectroscopic monitoring stage over $\sim2$ years, the flux of \hal\ varied minimally, whereas the fluxes of the Paschen lines changed significantly, by factors of up to $\sim2.3-3.7$. However, no clear evidence of kinematic variations was detected. 

    \item After considering the extinction correction, the Eddington ratio significantly increased from $\log \lambda_{\rm Edd} = -0.75$ to $-0.41$ and $-0.16$. This suggests that the accretion mode likely transitioned from a standard disk to a slim disk. 

    \item Based on the changes in the dust extinction and the Eddington ratio, we propose that we have witnessed structural changes in the dusty torus of NGC 3786, which served as the primary driver of the CL phenomenon. 

    \item The physical conditions within the BLR may be partially responsible for the changes in the flux ratios of the broad hydrogen emission lines. However, this is unlikely to be the sole cause of the CL phenomenon associated with the MIR-only outburst.  
   
\end{itemize}

\begin{acknowledgments}
We are grateful to the anonymous referee for the very constructive comments, which have greatly improved the manuscript. LCH was supported by the National Science Foundation of China (12233001) and the China Manned Space Program (CMS-CSST-2025-A09). S.S. was supported by the KIAA and Boya Fellow Grant of Peking University. This work was supported by the National Research Foundation of Korea (NRF) grant funded by the Korean government (MSIT) (Nos. RS-2024-00347548 and RS-2025-16066624) and the Yonsei University Research Fund of 2025 (2025-22-0402). D.K. acknowledges the support by the National Research Foundation of Korea (NRF) grant (No. 2021R1C1C1013580) funded by the Korean government (MSIT). This work was supported by K-GMT Science Program (PID: GN-2022B-Q-122 and GN-2023A-Q-125) of Korea Astronomy and Space Science Institute (KASI). 

\end{acknowledgments}

\bibliography{torus}

\appendix
\section{Temporal variations of Physical Properties}
Table~A1 lists the various properties of NGC 3786 obtained in different epochs. A detailed methodology for deriving the properties of NGC 3786 prior to the MIR outburst is presented in \citet{son_2022b}. The spectral fitting without kinematic constraints is shown in Figure~A1. The temporal variations of line fluxes in the follow-up period are shown in Figure~A2.  
\restartappendixnumbering

\begin{deluxetable*}{clrrrr}
\tablecolumns{6}
\tablenum{A1}
\tablewidth{0pc}
\tabletypesize{\scriptsize}
\tablecaption{Properties of NGC 3786 at Different Epochs\label{tab:table3}}
\tablehead{
\colhead{\h Property} &
\colhead{\h Data} &
\colhead{\h Before the MIR outburst} &
\multicolumn{3}{c}{After the MIR outburst} \\
\cline{4-6}
\colhead{\h} &
\colhead{\h} &
\colhead{\h} &
\colhead{\h} 2022-02-02&
\colhead{\h} 2023-06-10&
\colhead{\h} 2023-12-23\\
\colhead{\h (1)} &
\colhead{\h (2)} &
\colhead{\h (3)} &
\colhead{\h (4)} &
\colhead{\h (5)} &
\colhead{\h (6)}
}
\startdata
\hline
$E(B-V)$ &\h \hal/\pal &\h $\cdots$ &\h 0.92 &\h 1.42 &\h 1.60 \\
 &\h \hal/\pal(f) &\h $\cdots$ &\h $\cdots$ &\h 1.50 &\h 1.67 \\
\hline
$\log M_{\rm BH}$ &\h $L_{3.6\mu \rm m, bulge}$ &\h 6.70 &\h $\cdots$ &\h $\cdots$ &\h $\cdots$\\
\h &\h $\sigma_*$ &\h $6.83-7.54$ &\h $\cdots$ &\h $\cdots$ &\h $\cdots$\\
\h &\h $L_{{\rm H}\alpha}$ &\h 6.68 &\h 6.36 &\h6.33 &\h 6.37\\
\h &\h $L_{{\rm Pa}\beta}$  &\h $\cdots$ &\h 7.59&\h 7.18&\h 7.44\\
\h &\h $L_{{\rm H}\alpha}$ &\h $\cdots$ &\h 7.12&\h 7.52&\h 7.62\\
\h &\h $L_{{\rm H}\alpha}^{\rm c}$ &\h $\cdots$ &\h 6.84&\h 7.06&\h 7.20\\
\h &\h $L_{{\rm Pa}\beta}^{\rm c}$  &\h $\cdots$ &\h 7.72&\h 7.38&\h 7.68\\
\h &\h $L_{{\rm Pa}\alpha}^{\rm c}$ &\h $\cdots$ &\h 7.20&\h 7.64&\h 7.75\\
\hline
$\log L_{\rm bol}$  &\h $L_{[\rm O\, III]}$ &\h 43.70 &\h 44.04&\h $\cdots$ &\h $\cdots$\\
\h  &\h $L_{[\rm O\, III]}^{\rm c}$  &\h 43.14 &\h 43.44&\h $\cdots$ &\h $\cdots$\\
\h  &\h $L_{{\rm H}\alpha}$  &\h 42.66 &\h 43.31 &\h 43.24 &\h 43.34\\
\h  &\h $L_{{\rm H}\alpha}^{\rm c}$  &\h $\cdots$ &\h 44.05 &\h 44.39 &\h 44.64\\
\h  &\h $L_{{\rm Pa}\beta}$  &\h $\cdots$ &\h 43.96 &\h 43.84 &\h 44.31\\
\h  &\h $L_{{\rm Pa}\beta}^{\rm c}$  &\h $\cdots$ &\h 44.23 &\h 44.26 &\h 44.79\\
\h  &\h $L_{{\rm Pa}\alpha}$  &\h $\cdots$ &\h 43.91 &\h 44.18 &\h 44.40\\
\h  &\h $L_{{\rm Pa}\alpha}^{\rm c}$  &\h $\cdots$ &\h 44.05 &\h 44.39 &\h 44.64\\
\h  &\h $L_{{\rm H}\alpha}$(f)  &\h $\cdots$ &\h $\cdots$ &\h 43.23 &\h 43.34\\
\h  &\h $L_{{\rm H}\alpha}^{\rm c}$(f)  &\h $\cdots$ &\h $\cdots$ &\h 44.45 &\h 44.70\\
\h  &\h $L_{{\rm Pa}\beta}$(f)  &\h $\cdots$ &\h $\cdots$ &\h 43.82 &\h 44.34\\
\h  &\h $L_{{\rm Pa}\beta}^{\rm c}$(f)  &\h $\cdots$ &\h $\cdots$ &\h 44.26 &\h 44.84\\
\h  &\h $L_{{\rm Pa}\alpha}$(f)  &\h $\cdots$ &\h $\cdots$ &\h 44.23 &\h 44.45\\
\h  &\h $L_{{\rm Pa}\alpha}^{\rm c}$(f)  &\h $\cdots$ &\h $\cdots$ &\h 44.45 &\h 44.69\\
\h  &\h $M_{\rm W1}^{\rm a}$ &\h 42.88 &\h 43.35 &\h $\cdots$ &\h $\cdots$  \\
\h  &\h $M_{\rm W1}^{\rm b}$ &\h 43.47 &\h $\cdots$ &\h $\cdots$ &\h $\cdots$ \\
\hline
$\log (L_{\rm bol}/L_{\rm Edd})$ &\h $L_{[\rm O\, III]}$  &\h $-1.10$ &\h $-0.76$ &\h $\cdots$ &\h $\cdots$\\
\h  &\h $L_{[\rm O\, III]}^{\rm c}$ &\h $-1.66$ &\h $-1.36$ &\h $\cdots$ &\h $\cdots$\\
\h  &\h $L_{{\rm H}\alpha}$  &\h $-2.14$ &\h $-1.49$ &\h $-1.56$ &\h $-1.46$\\
\h  &\h $L_{{\rm H}\alpha}^{\rm c}$  &\h $\cdots$ &\h $-0.75$ &\h $-0.41$ &\h $-0.16$\\
\h  &\h $L_{{\rm Pa}\beta}$  &\h $\cdots$ &\h $-0.84$ &\h $-0.96$ &\h $-0.49$ \\
\h  &\h $L_{{\rm Pa}\beta}^{\rm c}$  &\h $\cdots$ &\h $-0.57$ &\h $-0.54$ &\h $-0.01$ \\
\h  &\h $L_{{\rm Pa}\alpha}$  &\h $\cdots$ &\h $-0.89$ &\h $-0.62$ &\h $-0.40$ \\
\h  &\h $L_{{\rm Pa}\alpha}^{\rm c}$  &\h $\cdots$ &\h $-0.75$ &\h $-0.41$ &\h $-0.16$ \\
\h  &\h $L_{{\rm H}\alpha}$(f)  &\h $\cdots$ &\h $\cdots$ &\h $-1.57$ &\h $-1.46$\\
\h  &\h $L_{{\rm H}\alpha}^{\rm c}$(f)  &\h $\cdots$ &\h $\cdots$ &\h $-0.35$ &\h $-0.10$\\
\h  &\h $L_{{\rm Pa}\beta}$(f)  &\h $\cdots$ &\h $\cdots$ &\h $-0.98$ &\h $-0.46$ \\
\h  &\h $L_{{\rm Pa}\beta}^{\rm c}$(f)  &\h $\cdots$ &\h $\cdots$ &\h $-0.54$ &\h $+0.04$ \\
\h  &\h $L_{{\rm Pa}\alpha}$(f)  &\h $\cdots$ &\h $\cdots$ &\h $-0.57$ &\h $-0.35$ \\
\h  &\h $L_{{\rm Pa}\alpha}^{\rm c}$(f)  &\h $\cdots$ &\h $\cdots$ &\h $-0.35$ &\h $-0.11$ \\
\h  &\h $M_{\rm W1}^{\rm a}$ &\h $-1.92$ &\h $-1.45$ &\h $\cdots$ &\h $\cdots$ \\
\h  &\h $M_{\rm W1}^{\rm b}$ &\h $-1.33$ &\h $\cdots$ &\h $\cdots$ &\h $\cdots$
\enddata
\tablecomments{
Col. (1): AGN property. 
Col. (2): Data used to estimate each property. ``f'' denotes the fitting result obtained without kinematical constraints on the broad emissions. 
Col. (3): Estimates based on the data obtained before the outburst.
Col. (4): Estimates based on the data obtained after the outburst.\\
$^{\rm a}$ W1 magnitude derived from the spectral energy distribution fit. \\
$^{\rm b}$ W1 magnitude derived from the 2D decomposition of the {\it Spitzer} 3.6 $\mu{\rm m}$ image. \\
$^{\rm c}$ Extinction-corrected. 
}
\end{deluxetable*}

\begin{figure*}[t!]
\centering
\includegraphics[width=0.9\textwidth]{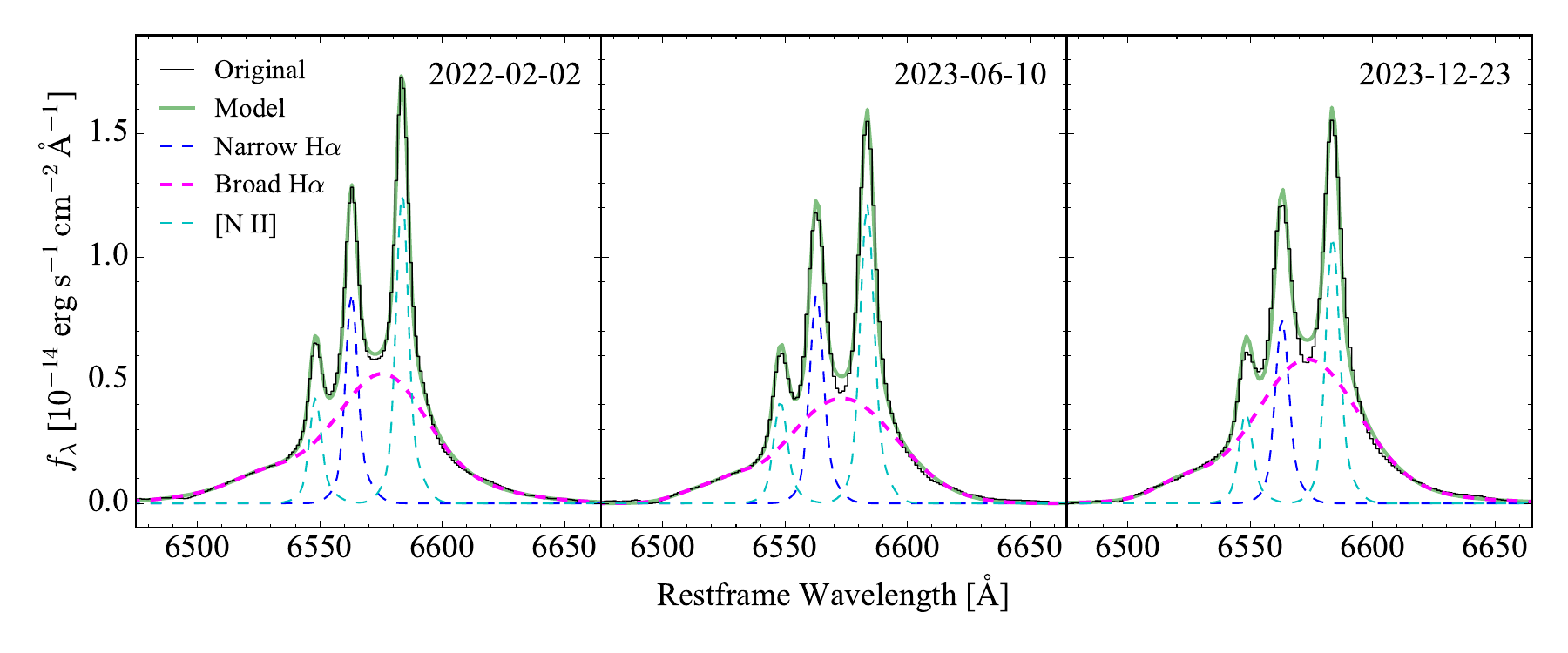}
\includegraphics[width=0.9\textwidth]{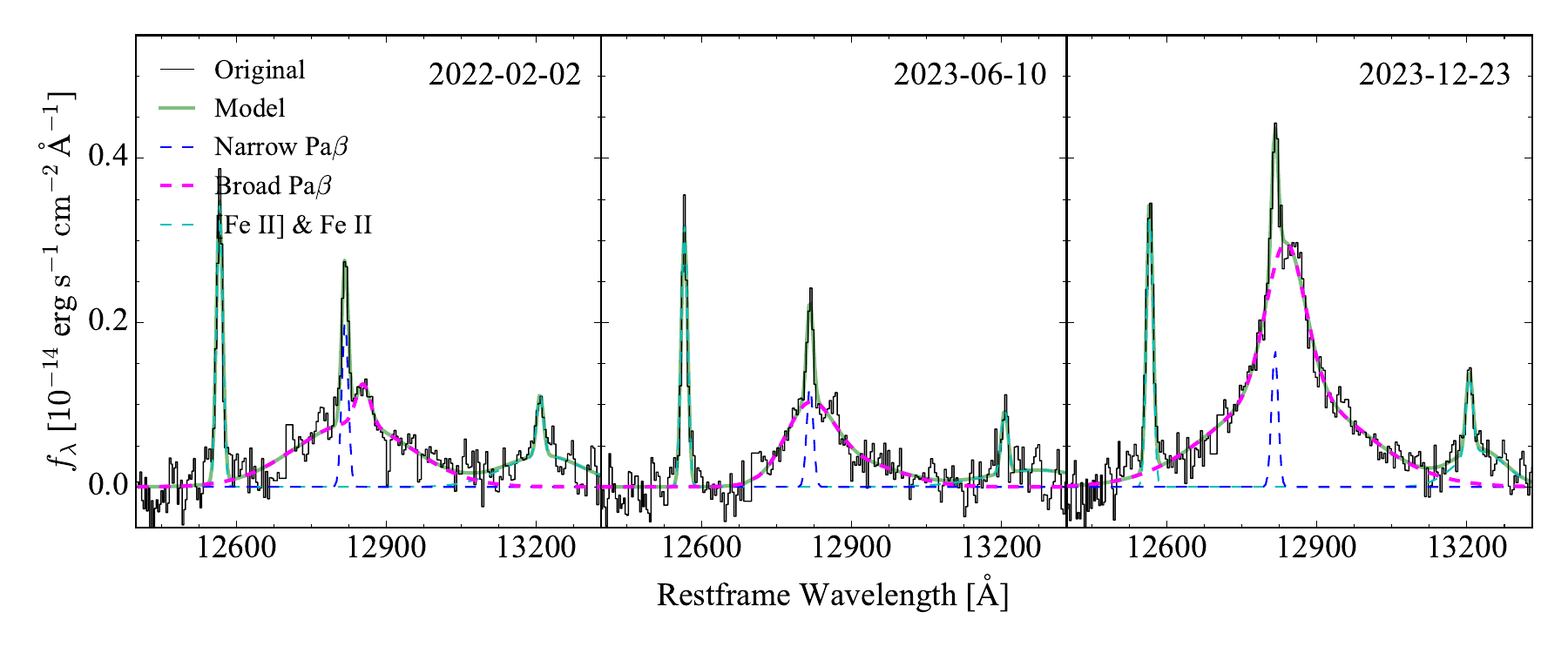}
\includegraphics[width=0.9\textwidth]{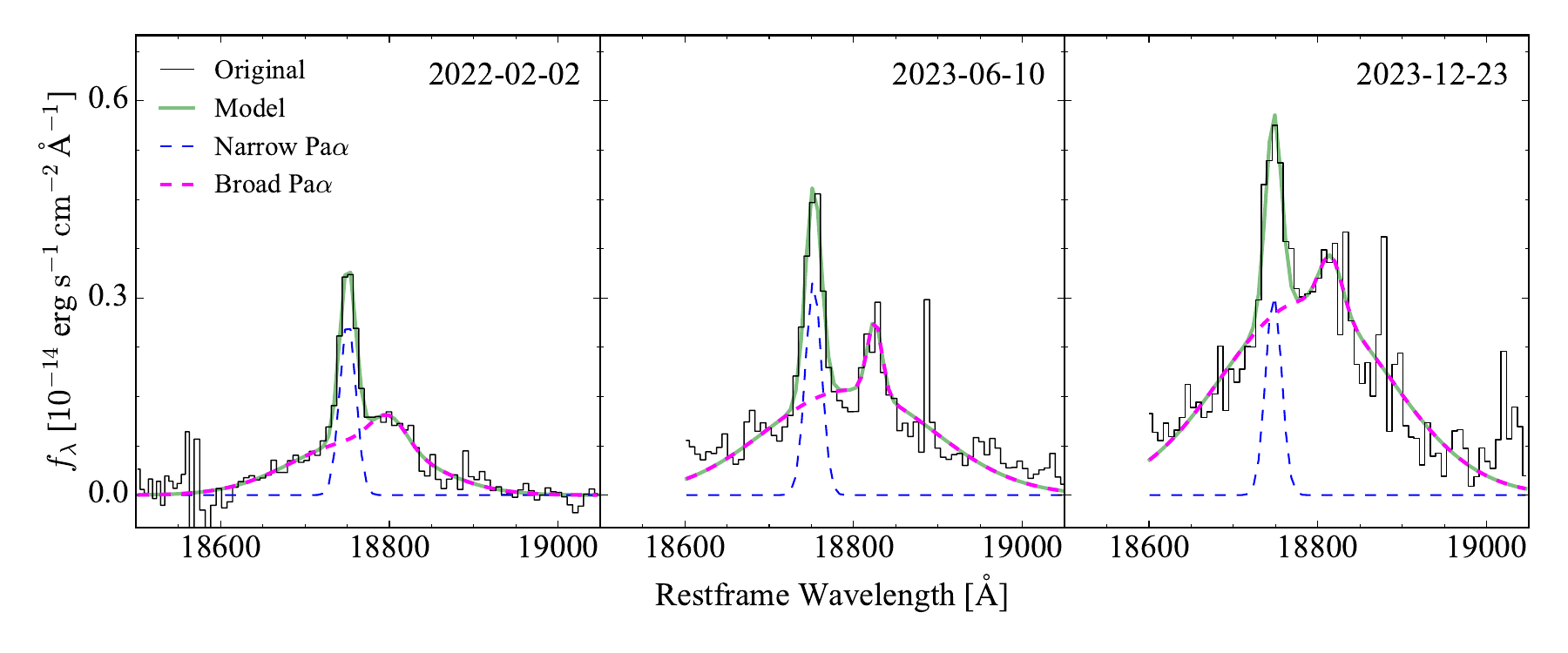}
\caption{
Same as Figure 2, but profiles of broad emission lines were allowed to vary freely during the fitting procedure.
\label{fig:fig4}}
\end{figure*}

\begin{figure}[ht!]
\centering
\includegraphics[width=0.5\textwidth]{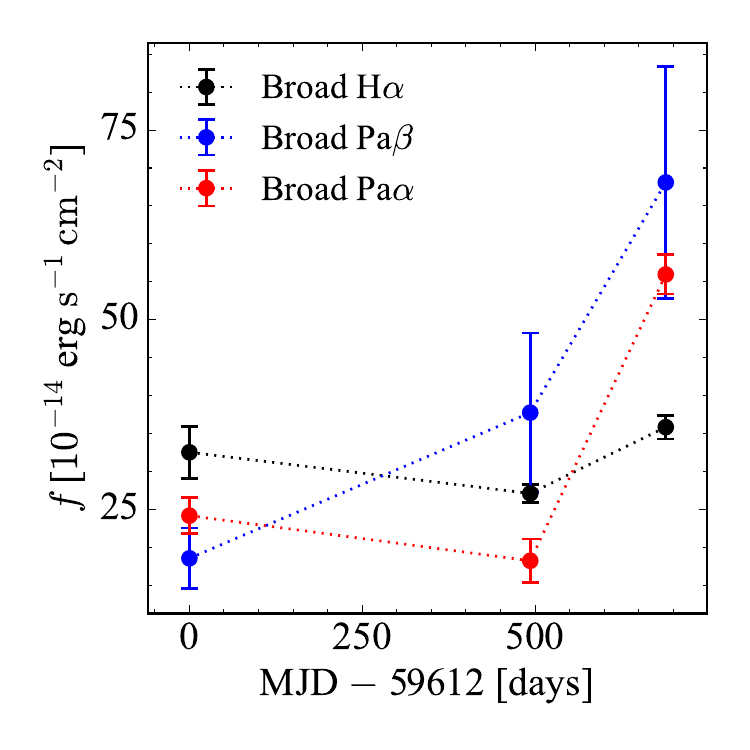}
\caption{
Temporal variations of fluxes for the broad \hal\, \pb, and \pal\ emission lines. 
\label{fig:fig5}}
\end{figure}

\end{document}